\documentclass[%
reprint,
showpacs,preprintnumbers,
 amsmath,amssymb,
 pre,
]{revtex4-1}

\usepackage{graphicx}
\usepackage{dcolumn}
\usepackage{bm}

\newcommand{\eref}[1]{Eq.~(\ref{eq:#1})}
\newcommand{\fref}[1]{Fig.~\ref{fig:#1}}
\newcommand{\sref}[1]{Sec.~\ref{sec:#1}}

\newcommand{\lr}[1]{ \left( #1 \right) }
\newcommand{\LR}[1]{ \left[ #1 \right] }
\newcommand{\llrr}[1]{ \left\{ #1 \right\} }
\newcommand{\av}[1]{ \left< #1 \right> }

\renewcommand{\vec}[1]{ \mathbf{#1} }

\newcommand{\position}{\vec{r}}
\newcommand{\velocity}{\dot{\position}}
\newcommand{\force}{\vec{F}}
\newcommand{\shearrate}{\dot{\gamma}}

\begin{document}

\preprint{APS/123-QED}

\title{Stochastic thermodynamics of a confined colloidal suspension under shear flow}

\author{Sascha Gerloff}
\email{s.gerloff@tu-berlin.de}
\author{Sabine H. L. Klapp}
\email{klapp@physik.tu-berlin.de}
\affiliation{%
 Institut f\"ur Theoretische Physik, Hardenbergstr. 36,\\ Technische Universit\"at Berlin, D-10623 Berlin, Germany
}%

\date{\today}

\begin{abstract}
Based on Brownian dynamics simulations, we investigate the thermodynamic signatures of non-equilibrium steady states in a confined colloidal suspensions under shear flow.
Specifically, we consider a thin film consisting of charged particles in narrow slit-pore confinement, forming two layers with quadratic in-plane structure in equilibrium.
This many-body system displays three distinct steady states, characterized by unique dynamical and rheological response to the applied shear flow.
Calculating the work and heat, we find that both quantities indicate the different states by their mean and by their distributions.
A particularly interesting situation occurs at large shear rates, where the particles perform a collective zig-zag motion.
Here, we find a bistability regarding the degree of phase synchronization of the particle motion.
It turns out that this bistability is key to understanding the resulting ensemble-averaged work distributions.
For all states, we compare the work and heat distributions to those of effective single-particle systems.
By this, we aim at identifying the many-body character of the stochastic thermodynamic quantities.
\end{abstract}
%
\maketitle
%
%
\section{\label{sec:introduction}Introduction}
In the past decades, the framework of stochastic thermodynamics (ST) \cite{Seifert2012,Sekimoto2010,VandenBroeck2015,Ciliberto2017} has attracted interest for a wide range of non-equilibrium systems, including molecular motors
\cite{Astumian1997,Wagoner2016,Seifert2011}, thermal conductors \cite{Ciliberto2013}, active matter \cite{Speck2016,Mandal2017,Gnesotto2018}, quantum systems
\cite{Campisi2011,Strasberg2013,Elouard2017} and colloidal systems \cite{Wang2002,Maggi2014,Rosinberg2017}.
One common feature of these systems is that they are subject to noise stemming from the coupling to a heat bath.
As a consequence, energy changes in the form of heat and work as well as the entropy production become fluctuating quantities.
These fluctuations can become especially important for small systems, such as molecular motors, which are largely driven by diffusion similar to a thermal ratchet \cite{Astumian1997}.
One important insight from ST is that these fluctuations can be restricted by fluctuation theorems, representing a generalized, stochastic counterpart of the second law of thermodynamics \cite{Seifert2012,Trepagnier2004,Crooks1999,Cohen2004}.

In the present study we apply concepts of ST to a driven colloidal suspension.
Indeed, in the context of ST, colloidal systems have proven to be a powerful "testing bed" in both, theoretical \cite{Seifert2012} and experimental investigations \cite{Ciliberto2017}.
Most of the research in this area has been concerned with single-particle systems, such as a single colloid driven by modulated \cite{Gieseler2015,Martinez2017} or translated optical trap \cite{Wang2002,Trepagnier2004}.
Further examples involve a colloidal particle driven over a periodic potential \cite{Gomez-Marin2006,Ma2017}, suspended in an active bath
\cite{Maggi2014,Mandal2017} or under delayed feedback control \cite{Rosinberg2017}, to name a few.
In contrast, systems containing many interacting particles have been less explored, although that one of the earlier reports of a FT was concerned with sheared molecular suspensions \cite{Evans1993}.
Only recently, concepts of ST are applied to systems with interacting degrees of freedom \cite{Ehrich2017}, examples being coupled oscillators \cite{Imparato2015}, driven colloidal monolayers \cite{Gomez-Solano2015}, soft particulate media in shear flow \cite{Rahbari2017}, and active matter \cite{Speck2016}.
A particularly interesting question for a many-particle system is how interaction-induced transitions between dynamical states are reflected in ST quantities.

In this study, we investigate, based on Brownian Dynamics (BD) simulations, a dense colloidal suspension confined to a narrow slit-pore, which is driven by a linear shear flow.
The interplay of narrow slit-pore confinement and shear flow induces rich non-linear dynamics, such as shear-induced melting and crystallization \cite{Derks2009, Vezirov2013}, buckling \cite{Cohen2004, Gerloff2017} as well as density heterogeneities \cite{Gerloff2016}.
Following previous studies by us \cite{Vezirov2013,Vezirov2015,Gerloff2016}, we focus on a system consisting of two layers with quadratic in-plane order in equilibrium, whose dynamical and structural response to the shear flow is characterized by three distinct steady states.
In the present study, we investigate the signatures of these steady states in two prominent (stochastic) thermodynamic quantities, that is, work and heat.
To this end, we study the mean values and distribution functions as functions of shear rate and time.
We also compare our numerical results with those of appropriate single-particle systems, which reveal many similarities particularly for small shear rates.
In contrast, at large shear rates, the work in the steady state is characterized by a bistability with regard to the degree of phase synchronization of the particle motion.
Here, the many-body nature of the system, which is reflected in the bistability, has a severe impact on the resulting work fluctuations.

The paper is structured as follows.
In \sref{planar slit-pore system} we briefly describe the model system and recapitulate the resulting dynamical behavior.
Section \ref{sec:stochastic thermodynamics} summarizes the concepts of stochastic thermodynamics of flowing systems \cite{Speck2008}, which we then apply to our sheared colloidal film in \sref{numerical results}.
In \sref{running state} we discuss, in particular, the bistability of the running state and its impact on the thermodynamics.
Finally, we compare our results to an effective single particle system in \sref{single particle on periodic potential} and conclude in \sref{conclusion}.
\section{\label{sec:planar slit-pore system} Planar slit-pore system}
\subsection{\label{sec:models and simulation details}Models and simulation details}
Following previous studies \cite{Vezirov2013, Vezirov2015, Gerloff2016}, we consider a colloidal suspension of charged macroions on a coarse-grained level, where the macroions interact via a combined \emph{Yukawa}-like- and soft-sphere (SS) potential.
The parameters are set in accordance to real suspensions of silica particles \cite{Klapp2007}.
The slitpore geometry is mimicked by two plane-parallel soft walls extended infinitely in $x$- and $y$-direction and separated in $z$-direction by a distance $L_z$.
The colloid-wall interaction is given by an integrated SS potential \cite{Jean-PierreHansen2013}.

We perform standard (overdamped) BD simulations to examine the nonequilibrium properties and dynamics of our model system.
The position $\position_i$ of particle $i$ is advanced according to the equation of motion \cite{Ermak1975}
\begin{equation}\label{eq:equation of motion}
 \velocity_i \lr{ t } =  \mu \;\force_i \lr{ \llrr{ \position } } + \dot{\vec{W}}_i\lr{t} + \shearrate z_i \vec{e}_x,
\end{equation}
where $\force_i$ is the total conservative force (stemming from two-particle- and the particle-wall interactions) acting on particle $i$ and $\llrr{ \position } = \position_1 ,\ldots, \position_N $ is the set of particle positions.
Within the framework of BD, the influence of the solvent is mimicked by a single-particle frictional- and random force.
The mobility is given as $\mu= D_0 /k_BT$, where $D_0$ is the short-time diffusion coefficient, $k_B$ is the Boltzmann constant, and $T$ is the temperature.
The random force is modeled by random Gaussian displacements $\delta \vec{W}_i$, with zero mean, variance $2D_0 \delta t$ for each Cartesian component, and the discrete time step $\delta t$.
The timescale of the system is set to $\tau=d^2/D_0$, which defines the so-called Brownian time.
We impose a linear shear profile $ \shearrate z_i \vec{e}_x $ [see last term in \eref{equation of motion}] representing flow in $x$- and gradient in $z$-direction.
The strength of the flow is characterized by the uniform shear rate $\shearrate$.
We note that, despite the application of a linear shear profile, the real, steady-state velocity profile resulting from the simulations can be non-linear \cite{Delhommelle2003, Gerloff2017}.

The present simulation approach has also been employed in other recent simulation studies of sheared colloids \cite{Besseling2012, Cerda2008, Lander2013};
the same holds for the fact that we neglect hydrodynamic interactions.
\subsection{\label{sec:shear-induced transitions in a bilayer system}Shear-induced transitions in a bilayer system}
\begin{figure}[t]
  \includegraphics[width=1.0\linewidth]{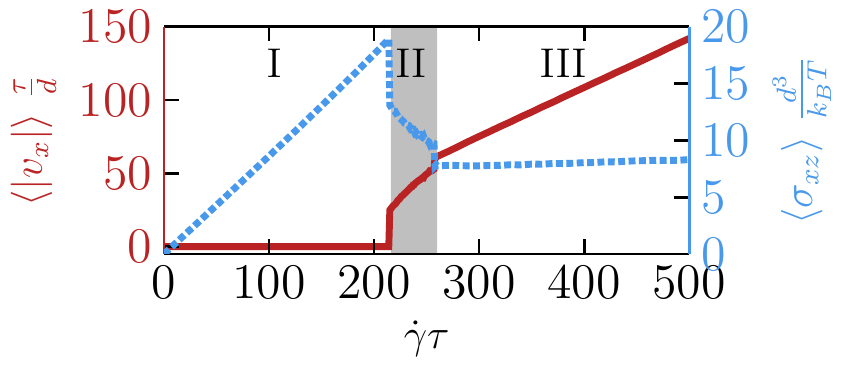}
  \caption{(Color online) Mean absolute velocity in $x$ direction, $\av{\left| v_x \right|}$, (solid red) and mean shear stress component, $\av{\sigma_{xz}}$, (dashed blue) as function of the dimensionless shear rate $\shearrate\tau$. The three distinct steady states are indicated by I (locked), II (disordered running) and III (ordered running). They correspond to a quadratic, disordered and hexagonal translational structure, respectively.}
  \label{fig:slitpore velocity and stress}
\end{figure}
Here, we recapitulate the well-explored dynamical behavior of a system which, in equilibrium, consists of two layers of colloidal particles with quadratic crystalline in-plane order \cite{Vezirov2013, Vezirov2015}.
Starting from equilibrium and applying a linear shear flow, the system displays three distinct steady states, as reflected in \fref{slitpore velocity and stress} by the nonlinear average absolute velocity $\av{\left| v_x \right|}$ and shear stress $\av{\sigma_{xz}}$ as a function of the dimensionless shear rate $\shearrate\tau$.
In the first (locked) steady state, the shear is insufficient to break the quadratic crystalline order, leading to elastic deformations while keeping the particles locked ($\av{\left| v_x \right|} = 0$).
Increasing the shear rate, the system undergoes a depinning transition to the second (disordered running) state at a critical shear rate $\shearrate_c\tau\approx 216$ \cite{Gerloff2016}, introducing net motion of the particles ($\av{\left| v_x \right|} > 0$) and breaking the crystalline in-plane order.
Another type of crystalline order is recovered for even larger shear rates ($\shearrate\tau\approx 260$), where the colloids reorganize into layers with hexagonal in-plane order, performing a collective zig-zag motion which characterizes the third (ordered running) steady state.
We note that the presence of a depinning transition, reflected by the sudden onset of nonzero average velocity, strongly reminds of the behavior observed in colloidal monolayers driven over a periodic substrate \cite{Gerloff2016}.
However, there is one conceptual difference: In our system there is no externally fixed substrate; rather the particles in one layer feel a fluctuating potential stemming from the other layer.

In the present study, we investigate the consequences of the nontrivial dynamical behavior of the sheared film from the point of view of (stochastic) thermodynamics.
\section{\label{sec:stochastic thermodynamics} Stochastic Thermodynamics}
Following Sekimoto's concept of stochastic energetics \cite{Sekimoto2010}, the terms appearing in the first law of thermodynamics,
\begin{equation}\label{eq:first law}
  dU = \delta w - \delta q\text{,}
\end{equation}
can be expressed on the basis of individual fluctuating particle trajectories.
Here, $\displaystyle dU$ is the total change of internal energy, $\displaystyle \delta w$ is the work applied to the system, and $\displaystyle \delta q$ is the heat dissipated \emph{into} the bath.
Considering, for simplicity, a (one-dimensional) overdamped stochastic process described by the Langevin equation $\displaystyle \dot{x} = -\mu \;\nabla_x U\lr{x,t} + \dot{W}$, the left hand side of \eref{first law} is given by
\begin{equation}\label{eq:total derivative potential energy}
  dU = \frac{\partial U}{\partial t} \circ dt - \frac{1}{\mu}\LR{ \dot{x} - \dot{W} }\circ dx \text{,}
\end{equation}
where, in the second term, we used that $\nabla_x U\lr{x,t} = -\mu^{-1}\LR{\dot{x}-\dot{W}}$, according to the Langevin equation.
In \eref{total derivative potential energy}, the symbol $\circ$ denotes the integration using the Stratonovich calculus.
The first term of \eref{total derivative potential energy} contains the temporal variation of the potential energy due to an external control protocol; it corresponds to the work $\delta w$ applied to the system.
The second term only involves contributions from the solvent, i.e. the friction forces $\mu^{-1}\dot{x}$ as well as the random forces $\mu^{-1}\dot{W}$; it can be identified as the heat $\delta q$ dissipated into the bath \cite{Sekimoto2010, Seifert2012}.
Note that both $\delta w$ as well as $\delta q$ are stochastic quantities with generally time-dependent distributions $P\lr{w}$ and $P\lr{q}$.
\subsection{\label{sec:some single particle results} Review of some single-particle results}
The extension of the expressions for the fluctuating work $w$ and heat $q$ to many-particle colloidal systems driven by an externally controlled potential is essentially straight forward.
However, finding exact solutions for these quantities and their distributions has, so far, only been accomplished for some single-particle systems (in presence of thermal noise).
One system of particular interest in the present study is a single particle in a moving harmonic trap $U\lr{x,t} = k/2\LR{ x\lr{t} - v\,t }^2$, whose center is translated with constant velocity $v$ \cite{Imparato2007, Wang2002, Seifert2012}.
We henceforth refer to this system as SP1.
In this system, the particle is "pinned" into the harmonic well for all driving speeds $v$, similar to a particle in a sheared colloidal layer within the locked (I) steady state (see \fref{slitpore velocity and stress}).
The probability distribution functions of the work $P\lr{w}$ and heat $P\lr{q}$ for SP1 can be calculated analytically \cite{Imparato2007}.
In particular, one finds in the steady state ($t\to\infty$) that the work is Gaussian distributed, with mean $\av{w} \approx \mu^{-1} v^2 t$ and variance $\av{\lr{w-\av{w}}^{2}} \approx  2 k_B T \av{w}$.
Note that, in this system, the variance and the mean are related via the integrated fluctuation theorem $\av{\exp\lr{-w}}=1$, which the work distribution obeys.
The heat distribution $P\lr{q}$ at $t\to\infty$ takes the form of a zeroth-order modified Bessel function of second kind in equilibrium ($v=0$) and becomes Gaussian distributed for $v > 0$ \cite{Imparato2007}.
In the latter case, the mean and the variance of $P\lr{q}$ are equal to that of the work distribution.

Many calculations performed for single particles in harmonic potentials can be generalized to rather arbitrary potentials \cite{Seifert2012}.
One important example is a single particle on a (one-dimensional) periodic substrate potential driven by a constant force \cite{Gomez-Marin2006,Seifert2012,Ma2017}.
For sufficiently large driving forces (or noise-levels), the particles are able to hop from one minimum to the next.
This situation somewhat resembles that encountered by a particle in the hexagonally ordered running (III) state of our sheared colloidal film.
In the single-particle case, the resulting distributions for the work $P\lr{w}$ and heat $P\lr{q}$ become non-Gaussian, yielding, for example, pronounced peaks in $P\lr{q}$ and a tilting of $P\lr{w}$ towards positive values for a particular choice of the substrate potential discussed in Ref. \cite{Gomez-Marin2006}.
In our study, one choice of particular interest is a single particle on a sinusiodal potential henceforth called SP2, which is translated with constant velocity $v$, corresponding to a driving due to a constant (flow) field $u=-v$.
Indeed, as we have shown in an earlier study \cite{Gerloff2016}, some aspects (particularly the depinning) of the sheared film can be understood by an effective one-dimensional model involving a sinusoidal potential.
To our knowledge, there are no analytical results for $P\lr{w}$ and $P\lr{q}$ in this case.
We therefore present and discuss in \sref{single particle on periodic potential} corresponding numerical results.

Yet another interesting example is a particle in a periodically modulated double-well potential \cite{Imparato2008, Fogedby2009}.
This system allows for hopping events between the two local minima of the double-well potential.
The corresponding work- $P\lr{w}$ and heat distributions $P\lr{q}$ are characterized by a peak at zero, as well as by pronounced shoulders.
In the low temperature limit, one finds for the heat distribution that these shoulders correspond to peaks at $q=\pm \Delta U$, where $\Delta U$ is the energy difference between the two minima, i.e. the energy difference after a hopping event \cite{Fogedby2009}.
\subsection{\label{sec:extension to many particle systems in flow}Extension to many particle systems in flow}
While the extension of ST to interacting many-body systems is straight forward, special care has to be taken when applying a flow field, as noted first in Ref.~\cite{Speck2008}.
The complications arise from the fact that the expressions for the work- and heat as introduced by Sekimoto are \emph{not} frame-invariant.
In particular, if the explicit time-dependence of the potential energy vanishes in the co-moving frame, the work $\delta w$ vanishes.
Therefore, the common expressions for the work- and heat [see \eref{total derivative potential energy} below] are only valid in the frame of reference where the solvent is at rest.

To account for the flow correctly, the authors of Ref.~\cite{Speck2008} proposed generalized expressions for the work- and heat rate, which are independent of the particular choice of the frame of reference.
The key idea is to add an advection term to the work rate and to measure the displacement of particles with respect to the flow.
The resulting work rate in a many-particle flowing system is given by
\begin{widetext}
  \begin{equation}\label{eq:work rate}
    \dot{w}\lr{t} = \sum_i \dot{w}_i\lr{t} = \sum_i \llrr{ \left.\frac{\partial U\lr{ \llrr{\position },t }}{\partial t} \right|_{\position_i} -  \vec{u}\lr{ \position_i, t }\cdot \vec{F}_i\lr{ \left\{ \position \right\}, t }
    + \vec{f}_i\cdot\LR{\velocity_i\lr{t} - \vec{u}\lr{\vec{r}_i,t} }  }\text{,}
  \end{equation}
\end{widetext}
where $\dot{w}_i\lr{t}$ is the work rate corresponding to particle $i$.
This expression takes into account three possible sources of work.
The first term in \eref{work rate} corresponds to the temporal variation of the potential energy $\displaystyle U\lr{ \llrr{ \position }, t }$ acting on particle $i$ due to an external control protocol.
The second term corresponds to the advection of particles due to an external flow with velocity $\displaystyle \vec{u}\lr{ \position_i ,t }$ against the conservative forces $\displaystyle \vec{F}_i\lr{ \left\{ \position \right\}, t } = -\nabla_i U\lr{ \llrr{ \position }, t }$.
The third term corresponds to the displacement of particles relative to the imposed external flow due to external (non-conservative) forces $\displaystyle \vec{f}_i$ acting on particle $i$.

The heat rate then follows from the first law of thermodynamics [see \eref{first law}] as
\begin{align}\label{eq:heat rate}
    \dot q\lr{t} &= \dot w\lr{t} - \frac{d\, U\lr{\left\{ \vec{r} \right\}, t} }{d t} = \sum_i \dot{q}_i\lr{t} \nonumber \\ &=\sum_i \LR{ \vec{f}_i + \vec{F}_i\lr{ \left\{ \vec{r} \right\}, t } } \cdot \LR{ \velocity_i\lr{t} - \vec{u}\lr{\vec{r}_i,t} }\text{,}
\end{align}
where $d\,U/dt = \partial\, U /\partial t + \sum_i \dot{\vec{r}}_i\lr{t}\cdot\nabla_i\, U$ denotes the total time derivative of the potential energy.
Note that, according to \eref{heat rate}, the heat dissipated into the bath vanishes whenever the particles follow the external flow perfectly, i.e. $\displaystyle \velocity_i\lr{t} = \vec{u}\lr{\position_i,t}$.
Integrating the work- and heat rates over time, we obtain the work and heat
\begin{equation}\label{eq:work}
  w\lr{t} = \sum_i w_i\lr{t} = \sum_i \int_0^{t} \dot{w}_i\lr{t'}dt'
\end{equation}
and
\begin{equation}\label{eq:heat}
  q\lr{t} = \sum_i q_i\lr{t} = \sum_i\int_0^{t} \dot{q}_i\lr{t'}dt'\text{,}
\end{equation}
respectively.
Here, the expressions in \eref{work rate}-(\ref{eq:heat}) were derived using the Stratonovich calculus.
Therefore, all derivates and integrals need to be performed using the mid-point rule, according to the Stratonovich interpretation (for more details see Appendix \ref{sec:app:stratonovich}).
\subsection{\label{sec:application to the sheared colloidal film under external flow} Application to the sheared colloidal film under external flow}
We now apply the expressions for the work- and heat rate to the sheared colloidal film.
To this end, we interpret the shear term in \eref{equation of motion} as an external flow, i.e. $\vec{u}_i = \shearrate z_i \vec{e}_x$.
Further, the external force $\vec{f}_i = 0$ is set to zero, since the system is solely driven by the shear flow.
Finally, we note that the total potential energy is not explicitly time-dependent.
The general expressions (\ref{eq:work rate}) and (\ref{eq:heat rate}) then simplify to
\begin{equation}\label{eq:work rate flow}
  \dot{w}\lr{t} = - \sum_i \vec{u}\lr{ \position_i }\cdot \vec{F}_i\lr{ \left\{ \position \right\} }\text{,}
\end{equation}
and
\begin{equation}\label{eq:heat rate flow}
  \dot q\lr{t} =\sum_i   \vec{F}_i\lr{ \left\{ \vec{r} \right\} }  \cdot \LR{ \velocity_i\lr{t} - \vec{u}\lr{\vec{r}_i} } \text{,}
\end{equation}
for the work- and the heat rate, respectively.

Inspecting \eref{work rate flow}, we find that the work rate is closely related to the instantaneous shear stress, which defines the mechanical response of the system to the applied shear.
This follows immediately if we insert the linear shear flow $\vec{u}_i = \shearrate z_i \vec{e}_x$ into \eref{work rate flow}, yielding
\begin{equation}\label{eq:work rate shear stress}
  \dot{w}\lr{t} = - \shearrate \sum_i z_i F_{x,i}\lr{ \llrr{\vec{r} } } = \shearrate \sigma_{xz} V\text{,}
\end{equation}
where $\sigma_{xz}$ is the x-z-component of the virial stress tensor $\bm{\sigma} = - V^{-1}\sum_i \position_i \vec{F}_i$, and $V$ is the volume of the slitpore.
This relation was also reported in other studies concerning shear driven systems \cite{Rahbari2017,Speck2017}.
Further, a similar relation was stated in an early study from Evans et al. \cite{Evans1993}, which proposes a linear relation between the heat used to thermostat the molecular dynamics simulations, and the shear stress.

From Eqs.~(\ref{eq:work rate flow}) and (\ref{eq:heat rate flow}), the time-dependent work and heat are obtained by numerical integration (see Appendix \ref{sec:app:stratonovich}) using Eqs.~(\ref{eq:work}) and (\ref{eq:heat}).

An alternative interpretation of the shear term in the equation of motion [see last term in \eref{equation of motion}] is presented in Appendix \ref{app:sec:external force interpretation}, where we interpret the shear as an external force.
This situation resembles to that discussed in a recent experimental study of colloidal monolayers driven over a periodic laser field \cite{Gomez-Solano2015}.
\section{\label{sec:numerical results}Numerical results}
\subsection{\label{sec:mean work and heat rate flow}Mean work and heat}
\begin{figure*}[t]
  \includegraphics[width=1.0\linewidth]{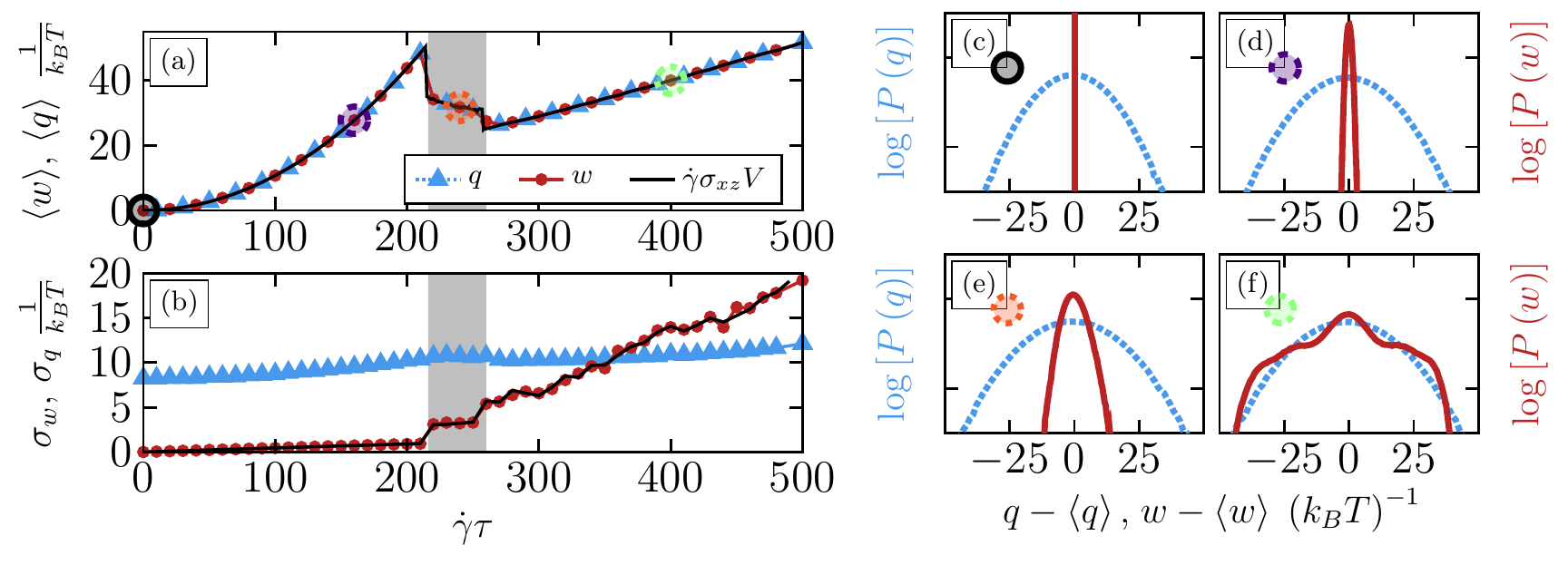}
  \caption{(Color online) (a) Mean value and (b) standard deviation of the work (red circles) and
  heat (blue triangles) related to the sheared colloidal film integrated over short time intervals ($\Delta t=10^{-5}\tau_B$), see \eref{work rate flow} and \eref{heat rate flow}, respectively.
  The mean is calculated by averaging over times $\tau_B$ and 100 \emph{independent} systems.
  The black solid line stem from the exact relation given in \eref{work rate shear stress}.
  (c-f) Corresponding distributions of the work (red) and heat (blue) for exemplary shear rates $\shearrate\tau=0,160,240,400$, respectively.
  These shear rates are marked in (a) with black, purple, orange and green circles and correspond to the equilibrium, locked (I)-, disordered running (II)- and ordered running (III) state.}
  \label{fig:short time work and heat flow}
\end{figure*}
To start with, we investigate the mean work $\av{w}$ and heat $\av{q}$ as a function of the shear rate, one main question being whether these quantities reflect the shear-induced transitions described in \sref{shear-induced transitions in a bilayer system}.
To this end we integrate the work $w$ and heat $q$ (before averaging) over short time intervals $\Delta t=10^{-5}\tau_B$.
In this limit, $w$ and $q$ are determined by the corresponding rates, e.g. $w \approx \dot{w}\Delta t$ and $q\approx\dot{q} \Delta t$.
The mean $\av{\cdot}$ is computed by analyzing trajectories for $100$ \emph{independent} systems at random times, yielding a combined ensemble and time average.
The results are shown in \fref{short time work and heat flow}(a).

We first investigate the work.
In equilibrium ($\shearrate\tau=0$), we find that $\av{w} = 0$, as expected since no work is performed on the system.
Applying a shear flow, $\av{w}$ displays a pronounced increase, reflecting that more and more work is consumed to maintain the quadratic crystalline structure characterizing the locked (I) steady state ($\shearrate\tau<216$).
Increasing the shear beyond the depinning threshold ($\shearrate\tau > 216$), $\av{w}$ jumps to smaller values and continues to decrease with the shear rate, corresponding to the disordered running (II) state, which appears upon breaking the quadratic crystalline order.
For shear rates $\shearrate\tau > 260$, $\av{w}$ performs another jump to smaller values and a subsequent increase, corresponding to the rearrangement of the colloids in hexagonal layers moving in a collective zig-zag motion [ordered running (III) state].
The overall behavior of $\av{w}$ is thus fully determined by that of $\av{\sigma_{xz}}$ \cite{Vezirov2015}, as expected from \eref{work rate shear stress}.

The mean heat (dissipated into the bath) displays the same behavior, not only from a qualitative point of view but also quantitatively.
Indeed, we find that $\av{w}=\av{q}$ for all considered shear rates, including the disordered running (II) state.
Such an equality of \emph{averaged} work and dissipated heat is expected in steady states, where the total time derivative of the energy should vanish on average.
While this is intuitively clear in the locked (I) and ordered running (III) state, the disordered running state (II) is more subtle:
Here, the negative slope of the shear stress suggests that this state is mechanically unstable and thus \emph{not} a true steady state for finite times.
In fact, state (II) is characterized by extremely long relaxation times \cite{Vezirov2015}.
Therefore, we think that there is indeed a very small difference between $\av{w}$ and $\av{q}$ in the disordered running (II) state.
This difference reflects the very slow relaxation dynamics, but it is too small to be detected in our simulations.
That is, on the time scales considered, the disordered running (II) state \emph{acts} like a steady state ($\av{w}\approx\av{q}$), while in reality it might be a transient state after all.
To confirm this point, further studies are necessary.

Overall, we find that the transitions between the different steady states (I-III) are clearly reflected in both, the mean work $\av{w}$ and heat $\av{q}$.
The corresponding shear rate dependence is closely related to that of the shear stress, as seen in \eref{work rate shear stress}.

\begin{figure}[t]
  \includegraphics[width=0.8\linewidth]{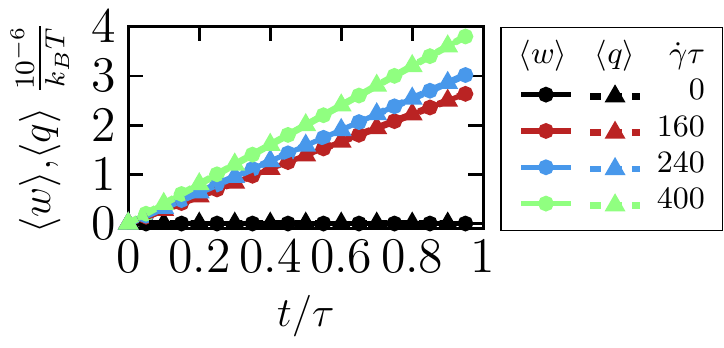}
  \caption{(Color online) Mean work $\av{w}$ and heat $\av{q}$ over integration time $t$ for four different shear rates $\shearrate\tau=0,160,240,400$ (black,red,blue,green).}
  \label{fig:mean work and heat time}
\end{figure}
Finally, in \fref{mean work and heat time}, we have plotted the mean work and heat as a function of the integration time $t$ for four shear rates $\shearrate\tau=0,160,240,400$ corresponding to the different steady states.
In equilibrium ($\shearrate\tau=0$), both the mean heat and work remain zero for all times, due to the absence of any driving forces.
Applying the linear shear flow, we observe a linear increase in time of the mean heat and work for all shear rates.
That is, the mean heat- and work \emph{rates} are constant, and \emph{on average} their shear rate dependency is given by the short-time heat and work, see \fref{short time work and heat flow}(a).
We note already here that we can understand the linear time-dependence as well as the quadratic increase of the mean work and heat in the locked (I) state by comparing to a well-studied single particle system (SP1), as discussed in \sref{comparison to single-particle systems}.
\subsection{\label{sec:work and heat distribution flow}Work- and heat distribution}
While the work- and heat are the same \emph{on average} $\av{w}=\av{q}$ for all steady states, this does not hold for the individual realizations of $w$ and $q$, which are subject to fluctuations.
To investigate these fluctuations, we consider the work- and heat distributions,  $P\lr{w}$ and $P\lr{q}$.
Results are plotted in \fref{short time work and heat flow}(c-f) for four exemplary shear rates $\shearrate\tau=0,160,240,400$, corresponding to equilibrium and the three steady states, respectively.
These results refer to a fixed, small integration time $t=10^{-5}\tau_B$.

Focusing first on $P\lr{w}$ and starting in equilibrium [$\shearrate\tau=0$, see \fref{short time work and heat flow}(c)], we find that $P\lr{w}$ is delta peaked.
This directly follows from \eref{work rate shear stress}, which shows that the work rate vanishes if no shear is applied to the system.
Applying a finite shear flow, the work turns out to be Gaussian distributed for all shear rates corresponding to the locked (I) state [$\shearrate\tau<216$, see \fref{short time work and heat flow}(d)].
That is, the skewness of the distribution
\begin{equation} \label{eq:skewness}
  \gamma_1^w=\frac{\av{\lr{w-\av{w}}^3}}{\sigma_w^3}\,\mathrm{,}
\end{equation}
vanishes and the kurtosis fulfills
\begin{equation} \label{eq:kurtosis}
  \gamma_2^w=\frac{\av{\lr{w-\av{w}}^4}}{\sigma_w^4} \approx 3\,\mathrm{,}
\end{equation}
where
\begin{equation} \label{eq:standard deviation}
  \sigma_w=\sqrt{\av{\lr{w-\av{w}}^2}}\,\mathrm{,}
\end{equation}
is the standard deviation.
Once the system melts [$\shearrate\tau>216$, see \fref{short time work and heat flow}(e)], $P\lr{w}$ becomes slightly skewed in positive direction with $\gamma_1^w \approx 0.225$.
Finally, for large shear rates [$\shearrate\tau>260$, see \fref{short time work and heat flow}(f)], $P\lr{w}$ displays pronounced asymmetric shoulders on both sides.
These stem from the collective hopping of the colloidal layers, which characterize the order running (III) state.
In fact, we find that the particular shape of $P\lr{w}$ is a result of a bistability, which we discuss in more detail in \sref{running state}.

For the heat distribution $P\lr{q}$, we find that, already in equilibrium, $P\lr{q}$ is Gaussian distributed, with vanishing skewness ($\gamma_1^q=0$) and kurtosis $\gamma_2^q=3$.
This remains true for all steady states.
In particular, we do not observe any shoulders in $P\lr{q}$ for the ordered running (III) state, contrary to the work distribution $P\lr{w}$.\\

We now turn to the dependence of the distribution on the integration time $t$.
In equilibrium, the work distribution remain delta peaked for all times, whereas the heat distribution quickly converges towards a static Gaussian distribution, as expected from the central limit theorem.
\begin{figure}[t]
  \includegraphics[width=1.0\linewidth]{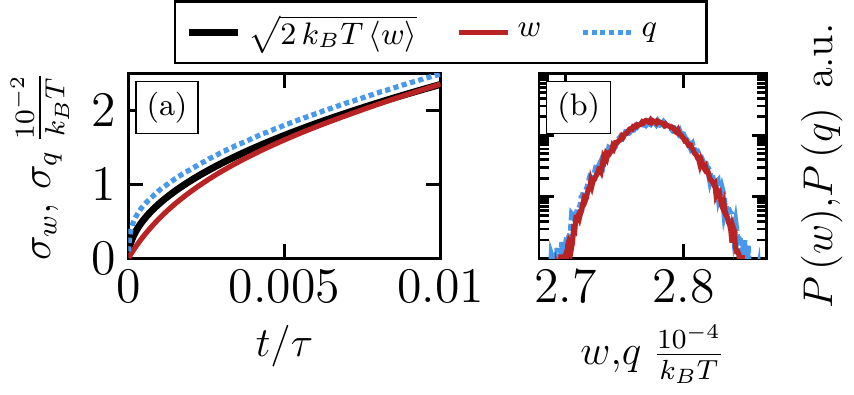}
  \caption{(Color online) (a) Standard deviation of the heat- (dashed, blue) and work distribution (solid, red) as a function of time for the locked (I) state $\shearrate\tau=160$. The square root of the mean value is plotted as a reference. (b) Corresponding distributions for integration times $t=0.01\tau$.}
  \label{fig:work and heat time locked}
\end{figure}
Applying a finite shear flow with $\shearrate\tau<216$ [locked (I) state], the work and heat distributions collapse over time to one Gaussian distribution, shown in \fref{work and heat time locked}(b) for $\shearrate\tau=160$.
While the mean of this distribution increases linear in time, the standard deviation $\sigma_{w/q}$ increases approximately with $\sqrt{2\,k_B T\av{w}}\propto \sqrt{t}$, plotted in \fref{work and heat time locked} (a).
This behavior is reminiscent of that found for particles trapped in harmonic traps, as we will discuss in more detail in \sref{comparison to single-particle systems}.

\begin{figure}[t]
  \includegraphics[width=1.0\linewidth]{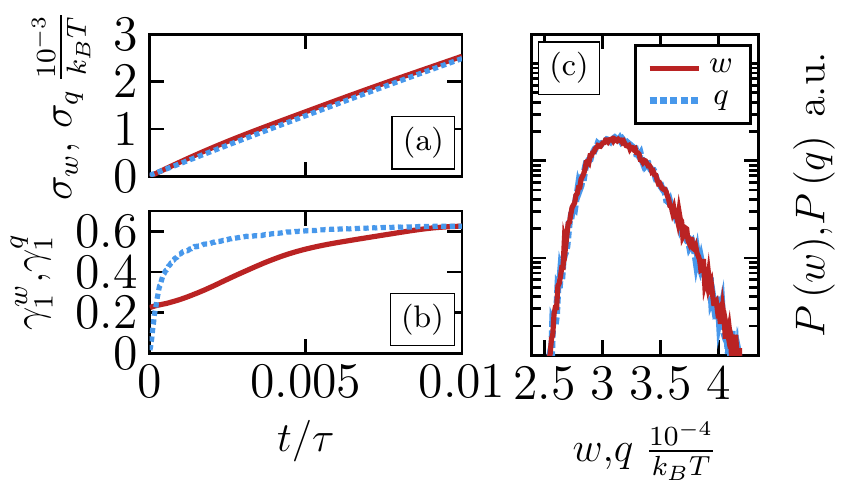}
  \caption{(Color online) (a) Standard deviation and (b) kurtosis of the heat- (dashed, blue) and work distribution (solid, red) as a function of time for the disordered running (II) state $\shearrate\tau=240$. (c) Corresponding distributions for integration times $t=0.01\tau$.}
  \label{fig:work and heat time melted}
\end{figure}
In the disordered running (II) state, the work and heat distributions again collapse onto a single distribution over time, as shown in \fref{work and heat time melted}(c) for $\shearrate\tau=240$.
For long integration times, the resulting distributions become strongly asymmetric, with the skewness saturating at $\gamma_1^{w/q}=0.625$, see \fref{work and heat time melted}(b).
In contrast to the locked (I) state, however, $\sigma_{w/q}$ now increases approximately linear with the mean $\propto \av{w}$, shown in \fref{work and heat time melted}(a).

Finally, in the ordered running (III) state, work and heat distributions display complex cyclic evolution, which we will discuss in \sref{work and heat distribution evolution ordered running state}.
\subsection{\label{sec:comparison to single-particle systems}Comparison to a single particle in a harmonic trap (SP1)}
In equilibrium ($\shearrate=0$), the particles are locked to a quadratic crystalline structure, where the position of each particle fluctuates around its lattice side.
For small fluctuations, we can approximate the potential energy experienced by one particle by a harmonic potential around its equilibrium position.
The work $P\lr{w}$ and heat distribution $P\lr{q}$ for such a single particle trapped in an effective harmonic trap can be calculated analytically \cite{Imparato2007}, as already briefly discussed in \sref{some single particle results}:
In equilibrium $P\lr{w}$ is given by a delta distribution at zero, while $P\lr{q}$ takes the form of a zeroth-order Bessel function of second type.
For the colloidal film, we also find a delta distribution for $P\lr{w}$ but the many-body heat distribution is Gaussian.
This is a result of the central limit theorem, which states that the distribution of the sum $q=\sum_{i=1}^N q_i$ of \emph{independent} and \emph{identical distributed} random variables $q_i$ will always converge towards a Gaussian distribution for large numbers $N\to\infty$ of summands.
Here, $q_i$ is the heat per particle.
The central limit theorem further predicts that the expected mean value is given by $E\LR{w} = N\, E\LR{w_i}$ and the standard deviation is $\sigma_N = \sqrt{N}\sigma_i$.
Evaluating the heat distribution per particle $P\lr{q_i}$ in equilibrium, we find that both relations hold and $P\lr{q_i}$ is indeed close to that of the expected Bessel function, consistent with the predictions of the single particle system.

At $\shearrate>0$, the locked (I) state is characterized by elastic deformations of the (quadratic) equilibrium structure.
Similar to the equilibrium, this situation is reminiscent of that of a single particle in a harmonic potential which is now translated with constant velocity $v$ through a resting solvent (SP1), inducing an effective solvent flow $u=-v$.
In SP1, the particle is pinned to the potential for all translation velocities $v$, which, due to the drag force, induces an elastic displacement of the particle from the trap center.
In the steady state, both the work and heat are Gaussian distributed \cite{Imparato2007}.
The mean work and heat $\av{w} = \av{q} \approx \mu^{-1} v^2 t$ are quadratic functions of the trap velocity $v$ and increase linear in time.
The standard deviation of the distributions, e.g. for the work $\sigma_w \approx \sqrt{ 2 k_B T \av{w}}$, is proportional to the square-root of the mean.
Comparing the flow velocities ($u$) from SP1 and the sheared colloidal film, we recognize that $v=-u\propto\shearrate$, which allows for a direct comparison.
Inspecting our results for the sheared colloidal film, we find that $P\lr{w}$ and $P\lr{q}$ are indeed approximately Gaussian [see \fref{short time work and heat flow}(d)] while the mean work $\av{w}$ and heat $\av{q}$ display a quadratic shear rate dependence [see \fref{short time work and heat flow}(a)] and a linear increase in time (see \fref{mean work and heat time}).
Further, the standard deviation of the work and heat is approximately given by $\sigma_{w/q} \approx \sqrt{2 \,k_B T \av{w}}$, consistent with SP1, for all shear rates corresponding to the locked (I) state [see \fref{work and heat time locked}(a)].
Note that in our sheared film, $\sigma_{q}$ displays a slight offset to larger values, whereas the standard deviation $\sigma_{w}$ is slightly smaller than $\sqrt{2 \,k_B T \av{w}}$ for small integration times.
These differences become negligible for large integration times $t\to\infty$.
Overall, we find that the results for $w$ and $q$ of the sheared colloidal film are in very good qualitative agreement with that of SP1.

Once the colloidal layers depin ($\shearrate\tau>216$) the comparison to SP1 fails, as expected since the particles are displaced far from their original interstitials.
In particular we observe that the work distribution becomes asymmetric for the disordered (II) running state [see \fref{short time work and heat flow}(e)] and displays pronounced shoulders in the ordered running (III) state [see \fref{short time work and heat flow}(f)].
We note that similar shoulders in the work distribution were observed for another system consisting of a single particle in a modulated double-well potential \cite{Imparato2008}.
In this system, these shoulders stem from the particle hopping from one minimum to the other, while the central peak corresponds to the particle staying at one particular minimum.
However, in the sheared colloidal film the mechanism is somewhat different and connected to a bistability of the ordered running (III) state, as discussed in the next section.
\section{\label{sec:running state}Ordered running state}
In part A of this section, we first discuss in some detail the particle motion in the ordered running (III) state.
Indeed, we have observed a new feature (not detected in our earlier studies), that is, a bistability.
This feature has strong impact on the work- and heat distributions, which we discuss in \sref{work and heat distribution evolution ordered running state}.
\subsection{\label{sec:bistability of running state}Microscopic motion}
\begin{figure}[t]
  \includegraphics[width=1.0\linewidth]{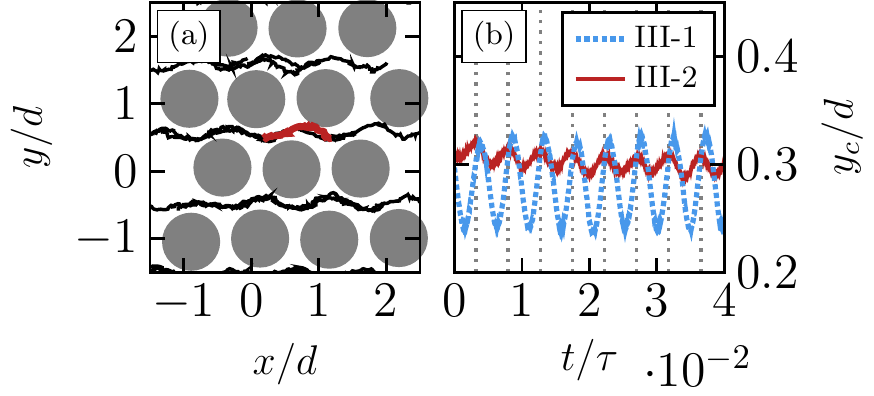}
  \caption{(Color online) (a) Segments of particle trajectories (black) of the top layer relative to the bottom layer (gray) at $\shearrate\tau=400$. A full zig-zag cycle of one particle is colored red for visibility. (b) Trajectories of the $y$-component of the center-of-mass of one layer for two different systems. The gray dotted lines indicate the period of the zig-zag motion, $\mathcal{T}=0.00476\tau_B$.}
  \label{fig:bistability running state}
\end{figure}
The ordered running (III) state is characterized by a hexagonal in-plane order and collective, oscillatory \emph{zig-zag} motion of the particles of each layer.
Specifically, the particle motion consists of periodic collective hopping from one interstitial of the neighboring layer to the next, as shown by the particle trajectories in \fref{bistability running state}(a).
Note that one full zig-zag cycle consists of two hopping events, where one is to the right ($y-\av{y}<0$) and the other to the left ($y-\av{y}>0$), with $\av{y}$ being the average position.

The oscillatory motion is also reflected by the $y$-component of the center-of-mass trajectories of the layers, see \fref{bistability running state}(b).
As discussed in a previous study \cite{Vezirov2013}, the frequency of these center-of-mass oscillations $\omega_0$ is related to the mean velocity $\av{\dot{x}}$ and the periodicity in flow direction, $x_0$, of the hexagonal in-plane structure, according to $\omega_0 = 2\pi \shearrate_\mathrm{eff}\Delta z/x_0$.
Here, $\shearrate_\mathrm{eff}$ is the effective shear rate extracted from the velocity profiles in $z$ direction $\dot{x}\lr{z} \approx \shearrate_\mathrm{eff}z$, for details see Ref. \cite{Vezirov2013}.
The length scale $x_0$ remains approximately constant while $\av{\dot{x}}$ increases linear with the shear rate (see \fref{slitpore velocity and stress}).
As a result, the oscillation frequency is fully determined by the shear rate.
Indeed, we find that the particle motion is synchronized with respect to their frequency for all shear rates corresponding to the ordered running (III) state.

However, a closer inspection of the center-of-mass trajectories of the individual realizations reveals that the amplitudes of the oscillations are not the same for all systems, in contrast to the frequency which remains constant.
In fact, we find that the systems separate into two sub-states, which are characterized by either large (III-1) or small amplitudes (III-2) of the center-of-mass oscillations.
In other words, the running state is characterized by a bistability.
Examples of the $y$-component of the center-of-mass trajectories for both sub-states are plotted in \fref{bistability running state}(b) for $\shearrate\tau=400$.
\begin{figure}[t]
  \includegraphics[width=1.0\linewidth]{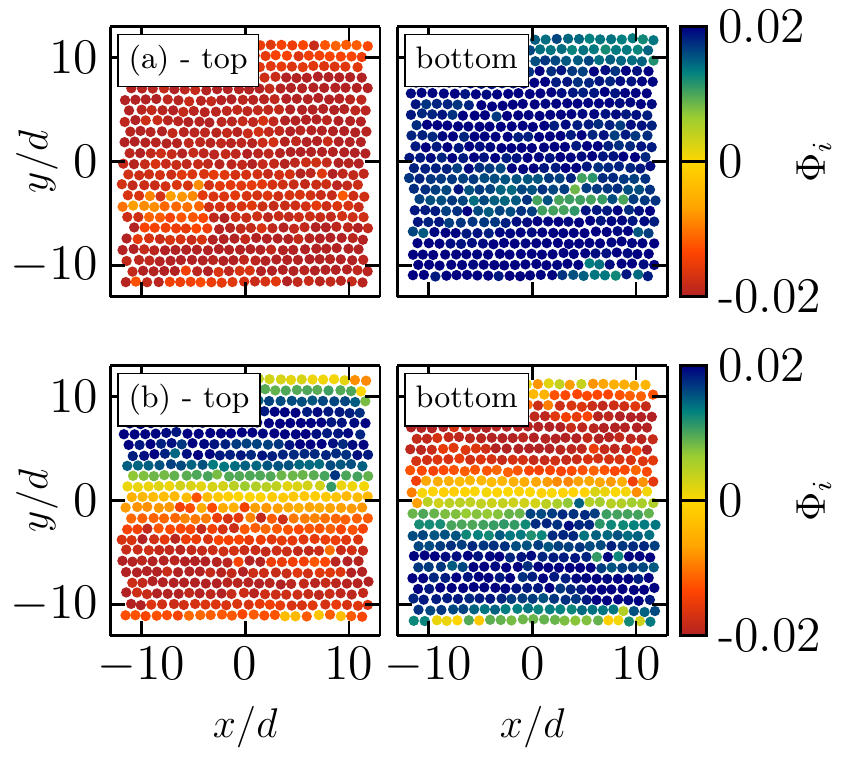}
  \caption{(Color online) Particle configurations inside the top (left) and bottom layer (right) for two different realizations corresponding to the in-phase (III-1) (a) and out-of-phase (III-2) sub-state (b) at shear rate $\shearrate\tau=400$. The color indicates the phase of the dominant frequency of the particle trajectories.}
  \label{fig:bistability phase per particle}
\end{figure}

To understand the nature of the two sub-states we analyze the trajectories of individual particles of one layer.
In particular, we calculate the Fourier transformation $\mathcal{F}_{y_i} \lr{ \omega_{0}}$ of the $y$-trajectories of each particle $i$, where the frequency $\omega_0 = 2\pi/\mathcal{T}$ corresponds to the mean period $\mathcal{T}$ of the zig-zag motion.
We note that $\omega_0$ corresponds to a maximum in the absolute value of the Fourier transform, as shown in Fig. 11 of Ref. \cite{Vezirov2013}.
Focusing now on the phase, i.e., $\Phi_i = \mathrm{Im} \LR{ \mathcal{F}_{y_i} \lr{ \omega_{0} } }$, we find that the spatial distribution of the phase is in general inhomogeneous and differs significantly for both sub-states (III-1) and (III-2).
This is visualized in \fref{bistability phase per particle}.

For the first (III-1) sub-state, we find that $\Phi_i$ is approximately constant within the layer, as shown in \fref{bistability phase per particle}(a) for the case $\shearrate\tau=400$.
This corresponds to the situation that the particle motion within the layer is fully synchronized with respect to their phase.
Note that the $y$-trajectories of the two layers are always in anti-phase to each other.
That is, when the particles of one layer jump to the left ($y-\av{y}>0$), the particles of the neighboring layer jump to the right ($y-\av{y}<0$), allowing for an efficient collective motion past each other in the presence of the slit-pore confinement.
This synchronized motion then results in large amplitudes of the center of mass motion of the layers [see \fref{bistability running state}(b)].

For the second (III-2) sub-state, we find \emph{two} domains with different phase $\Phi_i$, as shown in \fref{bistability phase per particle}(b) for a different realization at the same shear rate, $\shearrate\tau=400$.
That is, at the same time as some of the particles \emph{within} the layer perform a jump to the left ($y-\av{y}>0$) others perform a jump to the right ($y-\av{y}<0$).
As a result, the amplitudes of the center-of-mass oscillations are much smaller.
In fact, for a perfect anti-phase synchronized motion, of two domains of equal size, their oscillations would cancel each other perfectly, yielding a vanishing of the center-of-mass oscillations.
According to our observations, however, the amplitude remains finite, reflecting that the particle motion is not fully synchronized in an anti-phase manner.
We understand that this stems from fluctuations of the domain size and the resulting extended interfacial regions.
Inside these interfacial regions particle trajectories display a shift in phase such that they initiate hopping events at the same time as particles inside the two domains complete such events.
As a result, the interface regions are out of sync with the two anti-phase domains.

Given the observed bistability, it is an interesting questions to which extend this phenomenon depends on the size of the simulated system.
Running test simulations we find that the two states indeed exist for a wide range of system sizes.
Only for very small systems containing $N<200$ particles, the systems are always synchronized with respect to the phase.
For larger systems ($200<N<1058$), we find that the probability to find either sub-state (III-1 or III-2) seems to be approximately independent of the system size.
However, we expect that for very large systems ($N\to\infty$), the probability to find the fully synchronized state (III-1) should decrease significantly.
We also note, that the size of the domains is not constant for different system sizes.
Also, there are rare cases where the system spits into four domains rather than in two.
Overall we conclude, that the bistability is not a result of the particular system size chosen in our simulations.
\subsection{\label{sec:work and heat distribution evolution ordered running state} Work- and heat distributions}
\begin{figure}[t]
  \includegraphics[width=1.0\linewidth]{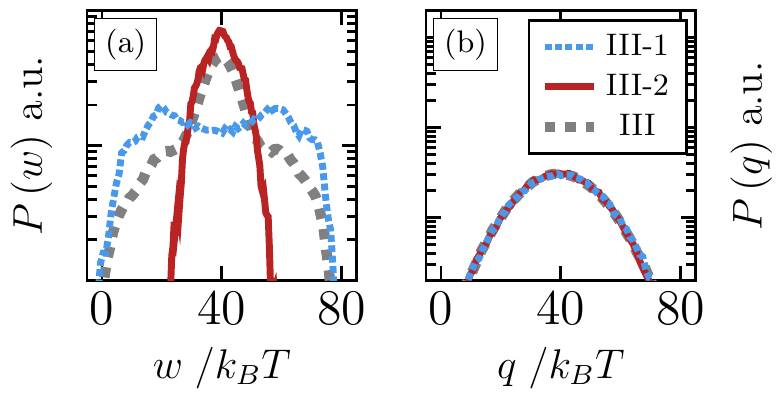}
  \caption{(Color online) (a) Work- and (b) heat distribution for the in-phase (III-1, blue) and out-of phase (III-2,red) sub-state at shear rate $\shearrate\tau=400$. The corresponding ensemble averaged distributions (gray, dashed) are plotted for reference.}
  \label{fig:bistability work distribution}
\end{figure}
With our findings concerning the bistability of the ordered running (III) state, we can now understand the ensemble-averaged work and heat in \fref{short time work and heat flow}(f) as a superposition of the individual distributions corresponding to the sub-states III-1 and III-2.
These distributions are plotted in \fref{bistability work distribution}.
While $P\lr{q}$ is the same, i.e. Gaussian, in both states [see \fref{bistability work distribution}(b)], $P\lr{w}$ for the two states is very different, as shown in \fref{bistability work distribution}(a).

In particular for the first, fully synchronized sub-state (III-1), the work distribution displays a pronounced double peaked structure, which can be related to the in-phase hopping of the particles.
Specifically, the right hand peak corresponds to the increased work required for particles to initiate a jump, whereas the left hand peak corresponds to the decreased work required for particles to complete a jump.
In fact, this situation is quite similar to that of a single particle driven on a periodic substrate (SP2), which we discuss in the next section.

For the second, partially synchronized sub-state (III-2), $P\lr{w}$ displays a rather narrow single peak.
We understand this as a result of the out-of phase motion:
Particles can initiate a jump at the same time as others complete one.
In such a situation, the width of the work distribution is decreased such that the two individual peaks corresponding to an in- or anti-phase motion can not be resolved.

Overall, we find that the III-1 sub-state contributes to the shoulders observed in the ensemble-averaged work distributions $P\lr{w}$ [see \fref{short time work and heat flow}(f)], whereas the the large central peak corresponds to the III-2 sub-state.\\
\\

\begin{figure}[t]
  \includegraphics[width=1.0\linewidth]{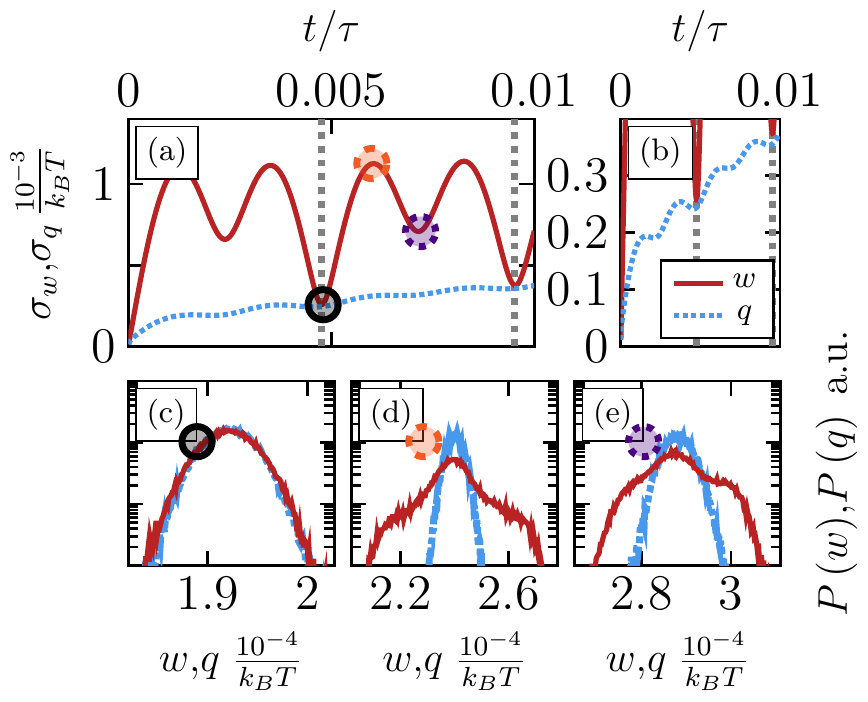}
  \caption{ (a) Standard deviation of the heat- (dotted, blue) and work distribution (solid,red) as a function of time for the ordered running (III) sub-state $\shearrate\tau=400$. The gray lines at $t=n\cdot0.00476\tau$ indicate the period of one full zig-zag cycle of the colloidal layers, shown in \fref{bistability running state}(b).
  (b) Enlarged view of the standard deviation of the heat.
  (c-e) Corresponding distributions for three integration times $t=0.00476, 0.006, 0.0072\tau$.}
  \label{fig:work and heat time running}
\end{figure}
We now turn to the time dependence of the work-  and heat distribution in the ordered running (III) state (averaged over the two sub-states).
Some results are shown in \fref{work and heat time running}(a-e).

In contrast to the locked (I) and the disordered running (II) state, $P\lr{w}$ and $P\lr{q}$ do \emph{not} collapse onto a single distribution for the considered integration times.
Instead, both evolve periodically in time, where the period $\mathcal{T}=0.00476\tau$ agrees with that of the zig-zag motion [see \fref{bistability running state}(b)].
The cyclic evolution, e.g. for  $\shearrate\tau=400$, is prominently reflected by the standard deviation $\sigma_{w/q}$ [see \fref{work and heat time running}(a) and (b)] of the corresponding distributions, which in turn are shown in \fref{work and heat time running}(c-e) for three different integration times $t=0.00476, 0.006, 0.0072\tau$.
In particular, both $\sigma_{w/q}$ display two local minima in one cycle, where the first minimum at $t=n\cdot\mathcal{T}$ is deeper than the second at $t\approx(n+0.5)\mathcal{T}$.
Closer inspection reveals that $P\lr{q}$ remains approximately Gaussian for all integration times, displaying only slight modulation of its width, as shown in the close-up in \fref{work and heat time running}(b).
In contrast, the shape of $P\lr{w}$ changes markedly in one period.
The periodic evolution of $P\lr{w}$ can be understood as follows.

At the beginning of a new cycle, at integration times $t=n\cdot \mathcal{T}$, $P\lr{w}$ conforms with $P\lr{q}$, i.e. collapse onto a Gaussian distribution, as shown in \fref{work and heat time running}(c) for $t=\mathcal{T}$.
Here, all particles have performed on average \emph{one} full zig-zag cycle.
For all realizations, the work consumed by this one cycle is approximately constant, leading to a narrow distribution of the integrated work.
In fact, the collapse of $P\lr{w}$ and $P\lr{q}$ suggests that the work at these integration times is determined mostly by the thermal fluctuations of the bath.

Upon increasing the integration time, $\sigma_w$ reaches an other local minimum at $t=0.0072\tau$ [see \fref{work and heat time running}(a)].
Here, all particles have performed a full zig-zag cycle as well as \emph{one} additional hopping of the next cycle.
Again, due to the hopping events consuming an approximately constant amount of work, the width of $P\lr{w}$ decreases.
However, at this time, $P\lr{w}$ and $P\lr{q}$ do not collapse, as shown in \fref{work and heat time running}(e).
Instead, $P\lr{w}$ displays pronounced asymmetric shoulders, stemming from the superposition of $P\lr{w}$ of the individual sub-states (III-1) and (III-2).
Similar to the short-time distribution [see \fref{bistability work distribution}(a)], systems in the latter sub-state contribute to the central peak and systems in the former contribute to the shoulders.
We note that this differs markedly from the distributions at integration times $t=n\cdot\mathcal{T}$ [see \fref{work and heat time running}(c)], suggesting that the two hopping events corresponding to one full cycle are not identical.

At times $t=0.006\tau$ and $t=0.0082\tau$, both $\sigma_{w/q}$ display two local maxima [see \fref{work and heat time running}(a) and (b)].
Here, the particles have performed, in addition to (at most) one jump of the current cycle, another segment of the full zig-zag motion.
These segments of the independent realizations are uncorrelated, leading to wide distributions of the heat and work.
At these times, $P\lr{w}$ displays rather wide symmetric shoulders [see \fref{work and heat time running}(d)], stemming again from systems in the (III-1) sub-state.

To better interpret the cyclic evolution of $P\lr{w}$, we consider in the next section again a single-particle system, namely a particle in a moving sinusoidal potential.
\section{\label{sec:single particle on periodic potential} Thermodynamics of an effective single-particle system}
In Ref. \cite{Gerloff2016}, we have developed a mapping strategy for the sheared colloidal film onto an effective one-dimensional model system, which estimates very well the location of the (depinning) transition from the locked (I) to the running (II-III) states of the sheared film.
This (deterministic) model consists of a single particle subject to an effective sinusoidal potential $V\lr{x} = V_0 \sin\lr{ 2\pi x / a_\mathrm{S} }$ driven by a constant flow $u$.
In absence of thermal noise, the equation of motion can be solved analytically.
For small flow velocities $u$, the particle is unable to overcome the potential barriers of the sinusoidal potential, corresponding to the locked (I) state of the sheared film.
Only for sufficiently large $u$, the particle is able to hop from one minimum to the next, corresponding to the ordered running (III) state.
As shown in Ref. \cite{Gerloff2016}, the model (adapted to the quadratic substrate of our film) yields a depinning threshold in very good agreement with the numerical data.
Moreover, it also captures key qualitative behavior for the running state.

In this section, we discuss the stochastic thermodynamics of the effective model in the presence of thermal noise.
The (overdamped) equation of motion reads
\begin{equation}\label{eq:equation of motion single particle on periodic potential}
  \dot{x} = \mu F_\mathrm{sub}\lr{x} + u + \dot{W}\lr{t}\text{,}
\end{equation}
where $\mu$ is the mobility, $F_\mathrm{sub}$ is the force stemming from the substrate potential $V\lr{x} = V_0 \sin\lr{ 2\pi x / a_\mathrm{S} }$, $u$ is the constant solvent flow velocity and $\delta W$ is a random Gaussian displacement.
Using the mapping strategy developed in \cite{Gerloff2016}, we identify the flow $u=\shearrate\av{\left| z \right|}$ of the single particle system with the corresponding $\shearrate$ of the slit-pore system.
The addition of thermal noise leads to thermally activated hopping events for shear rates below the depinning threshold $\shearrate\tau=216$.
This results in a continuous (rather than sharp) transition from the locked (I) to the running (III) state.
\subsection{\label{sec:effective single particle system numerical results}Work and heat distributions}
For the effective single particle system defined by \eref{equation of motion single particle on periodic potential}, the expressions for the work- and heat rate [see \eref{work rate} and \eref{heat rate} respectively] simplify significantly, yielding
\begin{align}
  \dot{w}\lr{t} &= - u\sum_i F_\mathrm{sub}\lr{x}\text{,} \label{eq:work rate single particle periodic potential}\\
  \dot q\lr{t} &=\sum_i   F_\mathrm{sub}\lr{x}  \cdot \LR{ \dot{x}_i\lr{t} - u } \label{eq:heat rate single particle periodic potential}\text{.}
\end{align}
The corresponding work and heat is calculated by simple integration according to \eref{work} and \eref{heat}.
We note, that the expressions
(\ref{eq:work rate single particle periodic potential}) and
(\ref{eq:heat rate single particle periodic potential}) are identical to that stemming from a particle subject to a sinusoidal potential that is translated with constant velocity $u$ relative to the solvent, i.e. $V\lr{x,t} = V_0 \sin\lr{ 2\pi (x+ut) / a_\mathrm{S} }$, where the work is given by
$\dot{w} = \partial V\lr{x,t}/\partial t$, yielding the same expression \eref{work rate single particle periodic potential}.
The same holds for the heat rate.

\begin{figure*}[t]
  \includegraphics[width=1.0\linewidth]{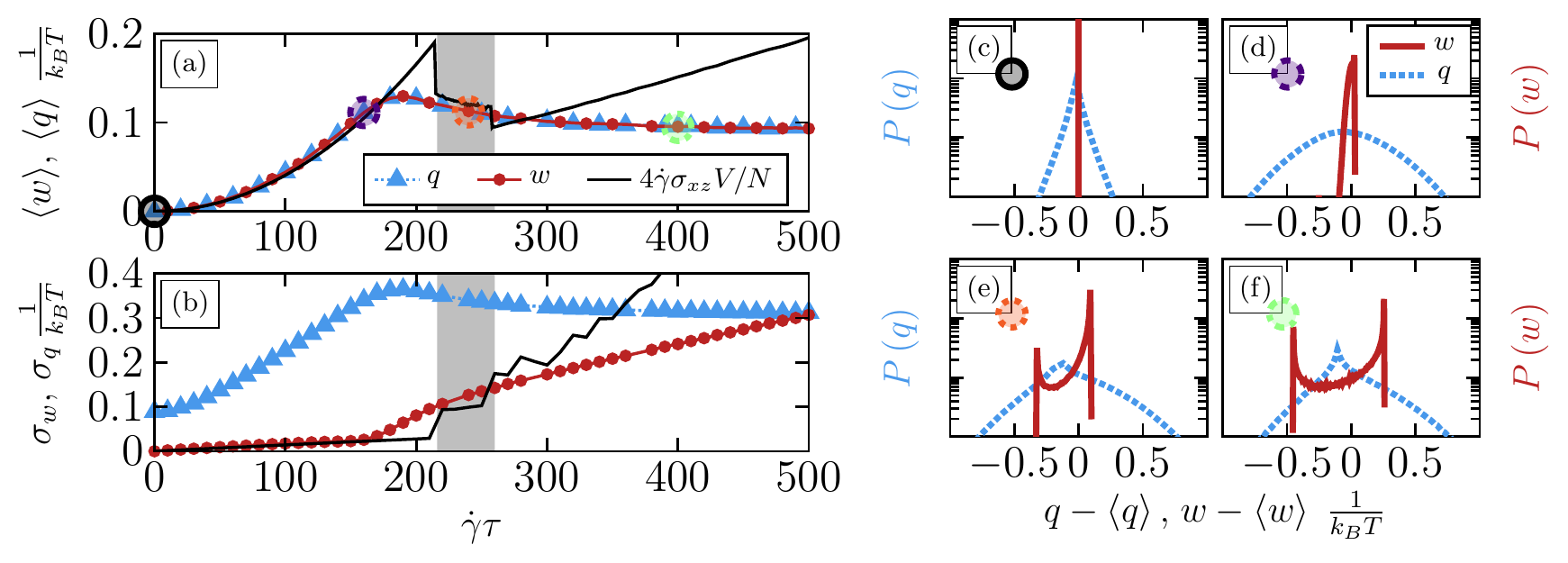}
  \caption{(Color online) (a) Mean value and (b) standard deviation of the work (red circles) and
  heat (blue triangles) integrated over short time intervals ($\Delta t=10^{-5}\tau_B$), see \eref{work rate single particle periodic potential} and \eref{heat rate single particle periodic potential}, respectively.
  The mean is calculated by averaging over times $\tau_B$ and 100 \emph{independent} systems.
  The black solid line stem from the estimation given in \eref{work rate shear stress} for the full system rescaled by the number of particles $N$ for the mean and $\sqrt{N}$ for the standard deviation respectively.
  (c-f) Corresponding distributions of the work (red) and heat (blue) for exemplary shear rates $\shearrate\tau=0,160,240,400$ (black, purple, orange, green circles), respectively.}
  \label{fig:short time work and heat flow single particle periodic potential}
\end{figure*}
Focusing first on the shear-dependency of the mean work $\av{w}$ and heat $\av{q}$ for short integration times $t=10^{-5}\tau$, plotted in \fref{short time work and heat flow single particle periodic potential}(a), we find that
$\av{w}=\av{q}$ are equal for all shear rates.
Starting from the equilibrium ($\shearrate=0$), where both the mean work and heat vanish, and applying a constant flow $u$, $\av{w}$ displays a pronounced quadratic increase for shear rates corresponding to the locked (I) state.
As discussed in \sref{comparison to single-particle systems}, the quadratic shear rate dependence stems from the elastic displacement of the particle from the (approximately harmonic) minimum of the substrate potential.
Once the particle depins from the substrate, the work and heat decrease and saturate for large shear rates corresponding to the running (III) state.
That is, for large shear rates, the impact of the substrate potential becomes negligible and driving the system by a constant flow requires on average a constant supply of work.

Turning now to the distributions functions, we find that, at $\shearrate = 0$, $P\lr{w}$ and $P\lr{q}$ agree with those for model SP1 (see \sref{comparison to single-particle systems}).
Applying the flow field $u=\shearrate\av{\left| z \right|}$, $P\lr{q}$ becomes Gaussian whereas $P\lr{w}$ displays a slight negative skewness ($\gamma_1^w > -0.5$), as shown in \fref{short time work and heat flow single particle periodic potential}(d).
Once the particle depins, $P\lr{w}$ transitions to a pronounced bimodal distribution displaying two sharp peaks [see \fref{short time work and heat flow single particle periodic potential}(e-f)], which correspond to the initiation and completion of a hopping event.
Here, the sharp bounding of $P\lr{w}$ can be understood from the equation of the work rate \eref{work rate single particle periodic potential}, which for a constant flow velocity $u$ is bounded by $\pm F_\mathrm{max} = V_0\;2\pi/a_\mathrm{S}$.
Note, that the right peak, corresponding to the particle initiating a jump, is much higher than the left, which corresponds to the relaxation towards the next minimum.
This stems from the fact that the build-up phase initiating a jump is much slower than the subsequent relaxation.
As a result, for large shear rates, the height difference between the left and right peak decreases, due to the jumping becoming more regular [see \fref{short time work and heat flow single particle periodic potential}(e-f)].
Finally, in this state, the heat distribution $P\lr{q}$ becomes non-Gaussian again, displaying two asymmetric flanks as well as a sharp peak near the middle.

\begin{figure}[t]
  \includegraphics[width=1.0\linewidth]{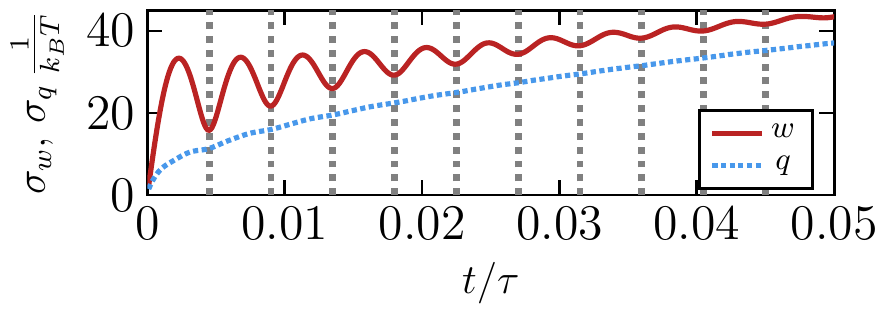}
  \caption{Standard deviation of the heat- (dashed, blue) and work distribution (solid,red) as a function of time for the running state $\shearrate\tau=400$ of the single particle system. The period of one hopping cycle $T=0.0045$ is indicated by dashed gray lines.}
  \label{fig:work and heat time 1D running}
\end{figure}
The work- $P\lr{w}$ and heat distributions $P\lr{q}$ depend not only on the shear rate, but also on the integration time $t$.
Here, we focus on the integration time dependence of $P\lr{w}$ and $P\lr{q}$ in the running state, which both display a cyclic time evolution (with that of $P\lr{w}$ being more pronounced than that of $P\lr{q}$), as shown in \fref{work and heat time 1D running}.
This cyclic evolution is characterized by local minima in the standard deviation of $P\lr{w}$ and $P\lr{q}$ at integration times $t=n\cdot \mathcal{T}$, corresponding to multiples of the mean period $\mathcal{T}=0.0045$ one particle takes to jump from one minimum to another.
At these times, all particles have performed one jump on average, which approximately consumes a fixed amount of work.
As a result, $P\lr{w}$ is mostly determined by the thermal fluctuations, thus leading to distributions similar to $P\lr{q}$.
In between these minima, the integration includes, in addition to a multiple of full jumps, an additional segment of the next jump.
Due to the independent choice of initial states, these segments are mostly uncorrelated, resulting in wide distributions of the work.
This cyclic evolution displays a pronounced damping, which reflect the fact that the oscillating particle motion becomes increasingly uncorrelated over time.
We contribute this to the thermal fluctuation driving the diffusion of the particles.
\subsection{\label{sec:effective single particle system comparison} Comparison to the sheared film}
Comparing the effective single particle system to the sheared colloidal film, we find many similarities.
In particular for the locked state, the single particle system gives quantitative predictions for the mean work and heat as a function of the shear rate.
Since the particle in the locked state is pinned to the (approximately parabolic) minimum of the periodic substrate potential, this situation can be already understood via SP1 (see \sref{comparison to single-particle systems}).

Here, we focus on the running state.
Comparing \fref{short time work and heat flow single particle periodic potential}(e-f) and \fref{bistability work distribution},
we realize that the work distributions $P\lr{w}$ of the single particle system indeed resembles that of the in-phase (III-1) running state of the sheared film:
The latter also displays two peaks corresponding to the initiation and completion of a jump.
The key difference is that for the single particle system, the two peaks are very sharp due to the boundedness of the substrate potential ($\pm F_\mathrm{max} = V_0\;2\pi/a_\mathrm{S}$), whereas the sheared film displays much wider peaks.
This stems from the fact that in the slit-pore, the "substrate" is a layer of particles which itself is subject to fluctuations.
As a result, the force required to initiate a jump is not constant, but a stochastic quantity itself, leading to wider peaks in $P\lr{w}$.

Another similarity is that $P\lr{w}$ evolves in a cyclic manner, where the period is determined by the average time $\mathcal{T}$ a particle needs to finish a jumping cycle.
For the single particle system, where each jump is identical, the time evolution is modulated by simple decaying oscillations.
In contrast, in the sheared film, each cycle displays two local minima as well as two maxima in between.
This more complex evolution stems from the fact that in the sheared film the particles need to perform two jumps in order to complete a cycle.
\section{\label{sec:conclusion}conclusion}
Using Brownian dynamics simulation we have investigated the stochastic thermodynamics of a thin colloidal film under planar shear flow.
Focusing on a particular parameter set corresponding to high density and strong spatial confinement, the colloids arrange, already in equilibrium, in two crystalline layers with quadratic in-plane order.
Applying the linear shear flow, the system displays three distinct steady states \cite{Vezirov2013, Vezirov2015, Gerloff2016}.
Using the framework of stochastic thermodynamics, we have calculated the work and heat of this many-particle system for all steady states, employing the expressions suggested in Ref. \cite{Speck2008}.
We find that, \emph{on average}, both the work and heat are related to the shear stress component of the configurational stress tensor, which is a common feature of shear driven systems \cite{Evans1993,Rahbari2017, Speck2017}.
Consistent with the shear stress, we therefore find jumps in the mean work and heat as a function of the shear rate, marking the borders of the domains of the steady states.
That is, all transitions between the three steady states are clearly reflected already in the \emph{mean} work and heat.
Moreover, the transitions are also reflected by the change of the shape of the work- and heat distributions, respectively, and their time evolution.

Particularly interesting distributions are found in the ordered running state, where the \emph{ensemble averaged} work distributions reflects a bistability regarding the degree of phase synchronization of the particle motion.
Here, one sub-state is characterized by global in-phase particle motions, whereas the other sub-state consists of two domains with opposite phase as well as extended interfacial regions.
The ensemble averaged work distribution is than a superposition of these sub-states, with a pronounced central peak and asymmetric shoulders.

To some extend, we can understand the work and heat distributions of the sheared colloidal film by comparing to appropriate single particles systems.
For the locked state, we find that the corresponding shear rate- and time-dependence of the work and heat distributions is in good agreement with that of a single particle trapped in a harmonic trap which is translated with constant velocity \cite{Imparato2007}.
For the ordered running state, in particular the in-phase sub-state, we can understand the work distributions by comparing to a single particle on a sinusiodal periodic potential driven by a constant flow.
In both cases, the work distributions display two peaks corresponding to the initiation and completion of hopping events, which evolve cyclically in time.
Overall, states where the individual particle motion is fully coherent are well described by effective single particle models for the center of mass.
The many-body character of the sheared film becomes apparent when the particle motion is disordered or only partially synchronized, as observed for the disordered running state and the out-of-phase sub-state.

Clearly, it would be very interesting to study the presence of fluctuation theorems for many-particle systems under shear.
There are several fluctuation theorems which predict certain relations for the distribution of the entropy production \cite{Seifert2012,Ehrich2017}.
In the steady state, this quantity is generally defined via the path probabilities of the particle trajectories \cite{Seifert2005}.
However, a derivation of these path probabilities for dense many-particle systems driven out of equilibrium is not trivial.
Therefore, one major challenge is to find appropriate expressions or methods to accurately compute the instantaneous entropy production from particle based simulations or experiments.
An other avenue would be to go to very large integration times.
In this limit, for an ergodic system, the system entropy production will vanish, leaving only the medium entropy production which is given by the heat \cite{Seifert2012}.
Therefore, one should recover the fluctuation theorems derived for the total entropy production for the heat distributions in the limit of long integration times, as it was done in Ref. \cite{Evans1993}.
However, since the heat is an extensive quantity, large system sizes and long integration times will shift the distributions towards large values.
This will substantially increase the computational effort to accurately sample the tails of the distribution.
To this end, focusing on smaller systems seems to be a promising strategy.
Work in this direction is in progress.

Related systems of particular interest are so-called nano-clutches, containing only few particles.
These are colloidal assemblies confined to a two dimensional circular geometry, which can be sheared by magnetic and optical torques \cite{Ortiz-Ambriz2018}.
These systems are experimentally accessible on the particle scale.
Here, it would be interesting to ask how the thermodynamic quantities are impacted by hydrodynamic interactions, as well as the different types of driving mechanisms.
In fact, in previous studies \cite{Vezirov2015} we have shown that the response of the system depends on the employed control scheme, i.e. shearing with constant rate or constant shear stress.
Thermodynamic quantities may serve as new control targets for control strategies, allowing to design the dynamical response of nano-machines.
At least from the simulation side, this might be a fruitful route for future studies.

\section{Acknowledgments}
This work was supported by the Deutsche Forschungsgemeinschaft through CRC 910: {\it Control of self-organizing nonlinear systems: Theoretical methods and concepts of application} (Project B2).
\appendix
\section{\label{sec:app:stratonovich} Stratonovich calculus}
It is well known that, when calculating the derivatives and integrals of stochastic quantities, extra care has to be taken \cite{Kampen1981} due to the fact that the rules of stochastic calculus are ambiguous.
In our numerical calculations we employ the Stratonovich calculus, consistent with the fact that this calculus was used to \emph{derive} the equations for the work- and heat rates [see \eref{work rate} and (\ref{eq:heat rate})].
To this end, the stochastic velocity $\velocity_i\lr{t}$ is determined using the mid-point rule
\begin{equation}\label{eq:numeric velocity}
  \velocity_i\lr{t} \approx \frac{ \position_i\lr{t+\Delta t} - \position_i\lr{t-\Delta t} }{ 2\Delta t }\text{,}
\end{equation}
where $\Delta t = 10^{-5}\tau_B$ is the time step of our BD simulations.
We note that using this definition is key to calculate the work- and heat rate given in \eref{work rate flow} and (\ref{eq:heat rate flow}).
In particular, the Stratonovich rule ensures that the correlation terms, such as $\av{ \force_i\lr{t}\cdot\position_i\lr{t\pm\Delta t} }$ arising in \eref{heat rate flow}, cancel properly.
Indeed, not using the mid-point rule may lead to unphysical results, such as a finite mean heat rate ($\av{\dot{q}}>0$) already in equilibrium.

Further, in order to calculate the work and heat we need to integrate the work- $\dot{w}_i\lr{t}$ and heat rate trajectories $\dot{q}_i\lr{t}$ numerically [see \eref{work} and (\ref{eq:heat})].
The corresponding integrals in the framework of the Stratonovich calculus, e.g. for the work, are given by
\begin{equation}\label{eq:numeric work}
  w_i\lr{t} = \int_{t_1=0}^{t_N=t}\dot{w}_i\lr{s} ds \approx \sum_{j=1}^{N-1} \frac{\dot{w}_i\lr{t_{j+1}} + \dot{w}_i\lr{t_j} }{2} \Delta t\text{,}
\end{equation}
where $t_{j+1} = t_j + \Delta t$ is a discrete series of times with $N$ entries and the step size $\Delta t = 10^{-5}\tau_B$.
The integral for the heat is evaluated in the same manner, respectively.
\section{\label{app:sec:external force interpretation}  External force interpretation}
In previous studies \cite{Gerloff2016}, we have shown that one can understand the transition from the locked (I) to the running (II-III) states in the sheared film as a depinning transition.
Therefore, our systems shares many similarities to driven monolayer systems, which consist of a single crystalline layer of colloids driven by constant force over an periodic substrate potential.
In a recent experimental study \cite{Gomez-Solano2015}, the stochastic thermodynamics of the latter system was investigated, reporting the work distributions for the pinned and running steady state.
They find, that in the pinned state the mean work done on the system vanishes, contrary to our findings according to which work is required to maintain the elastic deformations of the colloidal crystal layers.
In fact, this discrepancy is a result of an alternative interpretation of the driving mechanism, where one assumes an external force $f_i=\mu^{-1} \vec{u}\lr{\position_i}$ proportional to the flow field.

Applying this interpretation to our slitpore system, i.e. setting $\vec{u}_i = 0$ and $\vec{f}_i = \mu^{-1} \shearrate z_i \vec{e}_x$, the expressions for the work- \eref{work rate} and heat rates \eref{heat rate} reduce to
\begin{align}
  \dot{w}\lr{t} &= \sum_i \vec{f}_i\cdot \velocity_i\lr{t} \text{,} \label{eq:work rate external field}\\
  \dot q\lr{t} &=\sum_i \LR{ \vec{f}_i + \vec{F}_i\lr{ \left\{ \vec{r} \right\}, t } } \cdot \velocity_i\lr{t}  \label{eq:heat rate external field}\text{.}
\end{align}
Explicitly inserting the linear shear force $\vec{f}_i = \mu^{-1} \shearrate z_i \vec{e}_x$ into \eref{work rate external field}, yields further simplifications
\begin{equation}\label{eq:work rate mean velocity}
  \dot{w}\lr{t}  = \frac{\shearrate}{\mu}\sum_i  z_i \dot{x}_i \lr{t}  \approx \frac{\shearrate}{\mu} \sum_i  \av{z_i}_\text{EQ} \dot{x}_i \text{,}
\end{equation}
where $\av{z_i}_\text{EQ}$ is the mean position of particle $i$ in equilibrium.
In the last part of \eref{work rate mean velocity}, we used that due to the strong confinement, which restricts the $z$-position of the layers, the position of the individual particles is approximately constant such that $z_i \approx \av{z_i}_{EQ}$ .
Equation \ref{eq:work rate mean velocity} therefore shows that the work rate is approximately proportional to the velocity of the particles.
The corresponding work and heat are again obtained by integrating according to \eref{work} and (\ref{eq:heat}).

\begin{figure}[t]
  \includegraphics[width=1.0\linewidth]{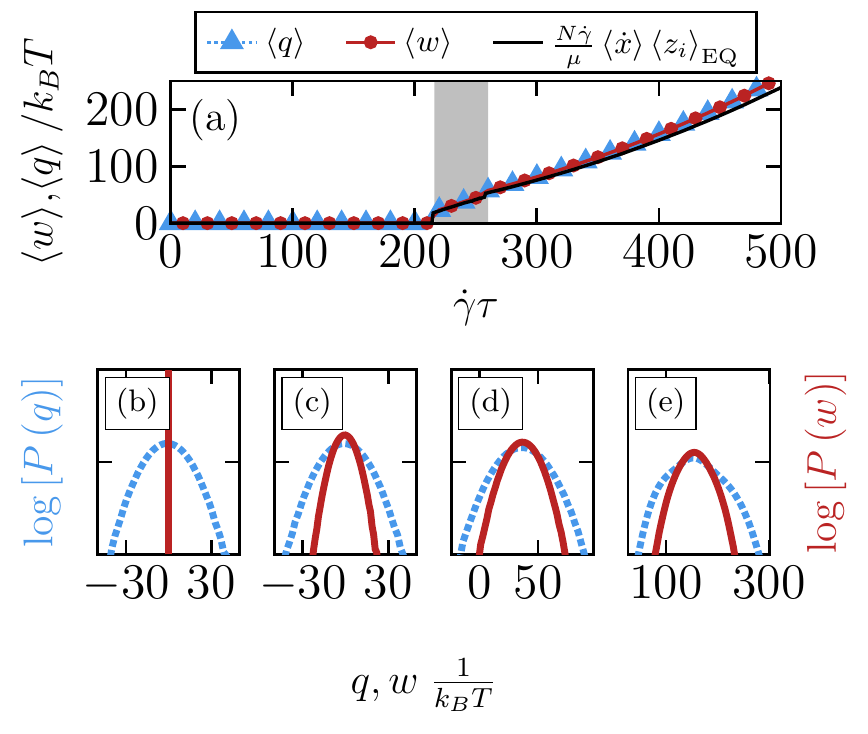}
  \caption{(Color online) (a) Mean short-time ($t=10^{-5}\tau_B$) work (orange circles) and heat (purple triangles) averaged over $\tau_B$ and 100 systems for the flow- [see \eref{work rate external field} and \eref{heat rate external field}]. The black dashed and solid line stem from the estimation given in \eref{work rate mean velocity}. (b-e) Corresponding distributions of the work (red) and heat (blue) for exemplary shear rates $\shearrate\tau=0,160,240,400$, respectively.}
  \label{fig:short time work and heat external force}
\end{figure}
In \fref{short time work and heat external force}(a), we have plotted the mean work $\av{w}$ and heat $\av{q}$ for short integration times $t=10^{-5}\tau_B$ using
\eref{work rate external field} and (\ref{eq:heat rate external field}).
Focusing on the work, starting in equilibrium, we find that $\av{w}$ again vanishes, as expected in absence of any driving forces.
However, applying a shear force, $\av{w}$ remains zero for the locked steady state ($\shearrate\tau<216$) and jumps to non-zero values only for shear rates corresponding to the unordered running state ($\shearrate\tau>216$).
For large shear rates ($\shearrate\tau>260$), corresponding to the ordered running state, $\av{w}$ displays a quadratic increase.
Again, we find that the domains of the different steady states are clearly reflected already in the mean work and heat, which are equal for all considered shear rates.

Examining now the distribution of the work $P\lr{w}$ [see \fref{short time work and heat external force}(b-e)], we find that $P\lr{w}$ is delta peaked in equilibrium and approximately Gaussian for all other steady states.
Only for longer integration times, $P\lr{w}$ displays asymmetric shoulders.
These results are consistent with the findings from Ref. \cite{Gomez-Solano2015}.
In our simulations we can further calculate the heat distributions, which are again Gaussian for the locked (I) and unordered running (II) state, see \fref{short time work and heat flow}(b-d).
Interestingly, in the ordered running (III) state ($\shearrate\tau>260$), the heat distribution displays pronounced asymmetric shoulders.

The discrepancy between the results in the flow (see \sref{numerical results}) and the force interpretation discussed in this section originate from the choice of frame of reference.
For the driven monolayer in its locked state, the particles are unable to follow the flow.
In the frame of reference of the solvent, the particles move on average with the flow velocity, requiring a continuous supply of work and corresponding to the flow interpretation.
In contrast, in the frame of reference of the substrate the particles are elastically displaced from their respective potential minima by a force proportional to the flow velocity, leaving the particles motionless on average.
This does not require any work and corresponds to the force interpretation.
Here, the work done by the flow on the particles is excluded from the system of interest and instead included in the energy of the bath, i.e the heat.

%

\end{document}